\documentclass[camera,letterpaper,nomarginnotes,nonarrowgutter]{jpaper}

\usepackage{cite}
\usepackage{amsmath,amssymb,amsfonts}
\usepackage{graphicx}
\usepackage{textcomp}
\usepackage{xcolor}
\usepackage{fancyhdr}
\usepackage[bookmarks=true,breaklinks=true,letterpaper=true,colorlinks,citecolor=blue,linkcolor=blue,urlcolor=blue]{hyperref}
\usepackage{tabularx}
\usepackage{multirow}
\usepackage{listings}
\usepackage{booktabs}
\usepackage{makecell}
\usepackage{enumitem}
\usepackage{balance}
\usepackage{datetime}
\usepackage{datetime2}
\usepackage{setspace}
\usepackage{adjustbox}
\usepackage{siunitx}
\usepackage{glossaries}
\usepackage[linesnumbered]{algorithm2e}
\setkeys{glslink}{hyper=false}

\newcommand{\x}{$\times$}                 
\newcommand{\degC}{\,$^{\circ}$C}

\def\BibTeX{{\rm B\kern-.05em{\sc i\kern-.025em b}\kern-.08em
    T\kern-.1667em\lower.7ex\hbox{E}\kern-.125emX}}
\begin{document}
\setstretch{0.987}
\title{Memory-Centric Computing: Security Benefits and Challenges\\of Processing-in-DRAM}

\author{{\.I}smail Emir Y{\"u}ksel\quad
F. Nisa Bostanc{\i}\quad
Ataberk Olgun\quad
Onur Mutlu
\vspace{2mm}
\vspace{-3mm}
\\\\
\emph{SAFARI Research Group}
\\
\emph{ETH Z\"urich}}
\maketitle

\thispagestyle{plain}
\pagestyle{plain} 
\pagenumbering{arabic}
\begin{abstract}
Today's computing systems are \emph{processor-centric}: they require frequent data movement between processing elements (e.g., CPU) and main memory (DRAM), leading to significant inefficiencies in performance and energy consumption. \emph{Memory-centric computing} instead moves computation to the data, enabling computation capability in and near all places where data is generated and stored, and greatly reducing the performance and energy overheads of data access and data movement. This shift from a processor-centric to a memory-centric paradigm has important and underexplored consequences for system security. Turning memory from a dumb, inactive store into an active computing substrate introduces benefits as well as challenges for system security: it can provide new in-memory security primitives and also reduce data exposure, but it can also expose new attack surfaces.

This work discusses the security benefits and challenges of memory-centric computing, specifically \emph{Processing-in-DRAM} (PiD), a paradigm where the operational characteristics
of a DRAM chip are exploited and enhanced to perform
computation on data stored in DRAM.  
Specifically, we describe 1)~new state-of-the-art DRAM-based true random number generators that provide up to 16.05~Gb/s throughput and physical unclonable functions with 5.75\% lower evaluation latency than the prior state-of-the-art, both on off-the-shelf DRAM chips and 2)~two key security challenges of PiD: amplified DRAM read disturbance (e.g., 158x reduction in the minimum number of DRAM accesses required to induce the first bitflip) and high throughput memory timing channels (e.g., a communication throughput of 14.8Mb/s).
We believe it is time to design, use, and program DRAM, and in general memory, not as an inactive storage substrate, but as a combined computation, storage, and security substrate, where computational capability, storage density, and security are all key goals.
\end{abstract}
\section{Security of Memory-Centric Computing}
\label{sec:intro}

Modern computing systems are \emph{processor-centric}. In this paradigm, memory is treated as a
dumb, inactive component: it serves the load and store requests of processing
units (e.g., CPU, GPU, TPU, FPGA, ASIC), but cannot operate on the data it holds or manage
itself. Computation can therefore happen \emph{only} after data moves across the memory
hierarchy to the processor. Such data movement is far more costly than computation,
in terms of energy, latency, and bandwidth~\cite{mutlu2019processing,mutlu2020modern,boroumand2018google,boroumand2021google,ghose.ibmjrd19}.
With the increasingly data-centric nature of contemporary and emerging applications such as generative artificial intelligence, large machine learning models, genome analysis, and video analytics, 
memory becomes an even larger performance, energy, robustness,
and system scaling bottleneck in processor-centric computing
systems~\cite{brown2020language,devlin2019bert,boroumand2021google,oliveira2022accelerating,heo2024neupims,zhou2022transpim,park2024attacc,seo2024ianus, rhyner2024analysis,yun2024duplex,jang2024smart,alser2020accelerating,singh2021fpga,alser2022from,kim2018grim,ghiasi2022genstore,ghiasi2024megis,cali2020genasm,cali2022segram,senolcali.bib2019,he2025papi,gu2025pim, ahn2015scalable,PEI,nai2017graphpim,besta2021sisa,salihoglu2013gps,tian2013from,low2012distributed,seshadri2017ambit,deoliviera2021damov,gomez2022benchmarking,boroumand2022polynesia}.

The \emph{memory-centric computing} (MCC) paradigm~\cite{ghose.ibmjrd19, mutlu2020modern, mutlu2019processing,mutlu2024memory,mutlu2025memory} can fundamentally solve data movement bottlenecks. 
The key idea is to place computation mechanisms in or near where the data is stored (i.e., inside the memory chips, in the logic layer of 3D-stacked memory, in the memory controllers, inside large caches, inside storage units or inside sensing units), so that data movement between where the computation is done and where the data is stored is reduced or eliminated, compared to contemporary processor-centric systems~\cite{HMC2, HBM, lee2016simultaneous, kim2024present, ahn2015scalable,PEI,hsieh2016transparent,boroumand2018google,boroumand2022polynesia,boroumand2021google,boroumand2019conda,singh2021fpga,oliveira2022accelerating,chi2016prime, Shafiee2016, seshadri2017ambit, seshadri2019dram, li2017drisa, seshadri2013rowclone, seshadri2016processing, deng2018dracc, xin2020elp2im, song2018graphr, song2017pipelayer,gao2019computedram, eckert2018neural, aga2017compute,dualitycache,besta2021sisa,seshadri2016buddy,seshadri.bookchapter17,seshadri2018rowclone,seshadri2015fast,li2016pinatubo,ferreira2022pluto,imani2019floatpim,he2020sparse,flashcosmos,truong2022adapting,truong2021racer,olgun2021quac,kim2019d,kim2018dram,bostanci2022dr,olgun2022pidram,ali2019memory,angizi2019graphide,li2018scope,subramaniyan2017parallel,zha2020hyper,fujiki2018memory,orosa2021codic,sharad2013ultra,rezaei2020nom,gao2021parabit,choi2020flash,han2019novel,merrikh2017high,wang2018three,lue2019optimal,kim2021behemoth,wang2022memcore,han2021flash,kang2021s,lee2020neuromorphic,lee20223d,si2019dual,
simon2020blade,nag2019gencache,wang2019bit,al2020towards,kang2014energy,kim2021colonnade,jiang2020c3sram,jeloka201628,wang2023infinity,kang2015energy,imani2020dual, chang2016low,hajinazar2021simdram,deng2019lacc,sutradhar2021look,sutradhar2020ppim,peng2023chopper,shahroodi2023swordfish,fernandez2024matsa,yuksel2024functionally,yuksel2025indram,yuksel2024simultaneous,tokuda2026pudghost,tokuda2026clutch,yang2026dcc}. 
MCC can be implemented in (i.e., using or near) different memory technologies~\cite{ghose.ibmjrd19, mutlu2020modern, mutlu2019processing}, including SRAM (e.g.,~\cite{aga2017compute,eckert2018neural,si2019dual,simon2020blade,nag2019gencache,wang2019bit,al2020towards,kang2014energy,kim2021colonnade,jiang2020c3sram,jeloka201628,wang2023infinity,kang2015energy,yuksel2022turan}), DRAM (e.g.,~\cite{ahn2015scalable,akin2014hamlet,akin2015data,ali2019memory,amiraliphd,angizi2019graphide,Asghari-Moghaddam_2016,asghari2016chameleon,azarkhish2016logic,azarkhish2018neurostream,babarinsa2015jafar,besta2021sisa,boroumand2018google, boroumand2019conda,boroumand2021google,boroumand2022polynesia,bostanci2022dr, cali2020genasm,CASES_MVX,chi2016prime,cho2020mcdram,C_RAM_1999,dai2018graphh,de2018design, deng2018dracc,devaux2019true,DRAMA_CAL_2014, drumond2017mondrian,farmahini2015nda,fernandez2020natsa,ferreira2022pluto,gao2016hrl,gao2017tetris,gao2019computedram,giannoula2022sparsep,gomez2021benchmarkingcut,gomez2022benchmarking,gu2020ipim,guo20143d,hsieh2016accelerating,hsieh2016transparent,huang2019active,huang2020heterogeneous,IRAM_Micro_1997,ke2021near,kersey2017lightweight,kim2018dram,kim2018grim,kim2019d,kwon202125,boroumand2016lazypim,lee2021hardware,li2017drisa,li2018scope,li2019pims,liu2018processing,nai2017graphpim, NDC_ISPASS_2014, NIM,niu2022184qps,olgun2021quac,olgun2022pidram,pattnaik2016scheduling, PEI, RVU,seshadri.bookchapter17, seshadri2013rowclone,seshadri2015fast,seshadri2016buddy, seshadri2016processing, seshadri2017ambit,seshadri2018rowclone, seshadri2019dram,shin2018mcdram,singh2019napel,singh2020nero,skhynixpim,Sparse_MM_LiM,sun2021abc,syncron,top-pim,tsai:micro:2018:ams, xin2020elp2im,Xi_2015,zhang2018graphp,zhuo2019graphq,lim2017triple,orosa2021codic,smc_sim,HIVE,jang2019charon,IBM_ActiveCube,rezaei2020nom,hall1999mapping,hadidi2017cairo,santos2018processing,MEMSYS_MVX,yuksel2024functionally,yuksel2025indram,yuksel2024simultaneous}), NAND flash (e.g,.~\cite{flashcosmos,gao2021parabit,choi2020flash,han2019novel,merrikh2017high,wang2018three,lue2019optimal,kim2021behemoth,wang2022memcore,han2021flash,kang2021s,lee2020neuromorphic,lee20223d,ghiasi2022genstore,jun2015bluedbm,ghiasi2024megis}), or emerging (e.g.,~\cite{fujiki2018memory,imani2019floatpim,Kim2016,li2016pinatubo,Shafiee2016,song2017pipelayer,song2018graphr,truong2021racer,zha2020hyper,truong2022adapting,sharad2013ultra,shahroodi2023swordfish}). We focus on DRAM~\cite{dennard1968field} due to its dominance as the main memory technology that can house large amounts of data. We broadly call MCC implemented in DRAM as \emph{Processing-in-DRAM} (PiD). PiD greatly reduces the performance and energy overheads of data access and data movement, and provides other benefits, e.g., improving system security and reducing system complexity. 

PiD can improve system security by providing new in-DRAM security primitives, reducing exposure of security-critical data, and accelerating security workloads (e.g., fully homomorphic encryption (FHE)~\cite{gburn2013homomorphic, tourky2016homomorphic, gentry2011implementing, al2020towards, gentry2009fully, van2010fully, cryptoeprint:2014/356, boneh2013private,fan2012somewhat}). A growing body of work demonstrates that real, unmodified commercial off-the-shelf DRAM chips can provide two key security primitives: by carefully violating DRAM access timing parameters and taking advantage of the resulting characteristics of different DRAM cells (i.e., whether they always/never fail or fail randomly),
it is possible to use DRAM to create physical unclonable functions (PUFs) and true random number generators (TRNGs) at high throughput and low latency~\cite{keller2014dynamic,sutar2018d,tehranipoor2016robust,eckert2017drng,kim2019d,talukder2019exploiting,pyo2009dram,olgun2021quac,bostanci2022dr,xiong2016run,schaller2019decay,liu2014trustworthy,zheng2021implementation,kumari2018rapid,mexis2025achieving,kim2018dram,talukder2019prelatpuf,hashemian2015robust,najafi2021deep,najafi2025epuf,miskelly2020fast,schaller2017intrinsic,fischer2025leveraging,anagnostopoulos2018intrinsic,li2023fphammer,orosa2021codic,gao2022frac,yuksel2025indram,baser2026simrapuf}, accelerating secure computation (FHE).

Unfortunately, PiD can also create new attack surfaces. Since PiD provides a direct, fast way to access and operate on memory, it can cause robustness (i.e., safety, security, reliability, availability) issues and unforeseen information leakage. Recent works demonstrate two such risks~\cite{yuksel2025pudhammer,bostanci2025revisiting}.First, Processing-using-DRAM (i.e., using DRAM cells to perform computation) relies on multiple-row activation, where multiple rows are activated either simultaneously or in quick succession, which can exacerbate DRAM read disturbance vulnerabilities~\cite{kim2014flipping,luo2023rowpress,yuksel2025columndisturb}.
Second, processing-in-memory provides direct main memory access to userspace applications such that they can bypass the cache hieararchy, 
which can be exploited to establish high throughput covert and side channels
through the shared DRAM row buffer~\cite{bostanci2025revisiting}.

In this paper, we describe examples from both security benefits and security challenges of Processing-in-DRAM. For security benefits, we cover 1) the state-of-the-art DRAM-based TRNG, SiMRA-TRNG~\cite{yuksel2025indram} and 2) the state-of-the-art DRAM-based PUF, SiMRA-PUF~\cite{baser2026simrapuf}. For security challenges, we cover 1) a recent experimental study on read disturbance effects of Processing-in-DRAM, PuDHammer~\cite{yuksel2025pudhammer}, and 2) a set of high throughput main memory-based timing attacks that exploit the characteristics of Processing-in-DRAM, IMPACT~\cite{bostanci2025revisiting}. 

We believe that the future of memory-centric systems is very bright and promising, yet
many exciting challenges remain to be solved across the computing stack to enable their \emph{secure,
widespread, and easy} adoption.

\section{Security Benefits of Processing-in-DRAM:\\In-DRAM Security Primitives}

Secure computation is of critical importance in modern computing systems. Therefore, it is important for a
memory-centric system to support fundamental security primitives
that enable secure computation and security functions.
Doing so would enable memory-centric systems to execute a wider
range of workloads securely. We focus on two key security primitives that DRAM provides: 1) true random number generation and 2) physical unclonable functions.
\subsection{In-DRAM True Random Number Generation}

Random number generators (RNGs) are critical components in many different applications, including cryptography, scientific simulation, industrial testing, and recreational entertainment~\cite{rock2005pseudorandom, ma2016quantum, stipvcevic2014true,
barangi2016straintronics, tao2017tvl, gutterman2006analysis, von2007dual,
kim2017nano, drutarovsky2007robust, kwok2006fpga, cherkaoui2013very,
zhang2017high, quintessence2015white}. In particular, for modern cryptographic applications, a random  number generator is critical to prevent information leakage to a potential adversary~\cite{kocc2009cryptographic, gutterman2006analysis,
von2007dual, kim2017nano, drutarovsky2007robust, kwok2006fpga,
cherkaoui2013very, zhang2017high, quintessence2015white}. 
RNGs are broadly classified into two categories~\cite{tilborgencyclopedia, chevalier1974random, knuth1998art,
tsoi2003compact}: pseudo-random number generators (PRNGs)~\cite{matsumoto1998mersenne, blum1986simple, mascagni2000algorithm,
steele2014fast, marsaglia2003xorshift}, which deterministically generate numbers starting from a seed value to approximate a true random sequence, and true random number generators (TRNGs)~\cite{hashemian2015robust,
 wang2012flash, ray2018true, holcomb2007initial,
holcomb2009power, van2012efficient, chan2011true, tzeng2008parallel,
teh2015gpus, majzoobi2011fpga, wieczorek2014fpga, chu1999design,
amaki2015oscillator, mathew20122, brederlow2006low, tokunaga2008true,
bucci2003high, bhargava2015robust, kinniment2002design, holleman20083,
gutterman2006analysis, dorrendorf2007cryptanalysis, lacharme2012linux,
pareschi2006fast, yang2016all,keller2014dynamic,sutar2018d,tehranipoor2016robust,eckert2017drng,kim2019d,talukder2019exploiting,pyo2009dram,olgun2021quac,bostanci2022dr,yuksel2025indram}, which generate random numbers by sampling non-deterministic physical phenomena. An effective TRNG must 1)~produce truly random numbers, 2)~provide a high throughput of random numbers at low latency, and 3)~be practically implementable at low cost.

DRAM offers a promising substrate for developing an effective and widely-available TRNG due to the prevalence of DRAM throughout all modern computing systems, ranging from microcontrollers to supercomputers. 
A high throughput DRAM-based TRNG would help enable widespread adoption of applications that are today limited to only select architectures equipped with dedicated high-performance TRNG engines. In addition to traditional computing paradigms, DRAM-based TRNGs can benefit memory-centric systems: a low-latency, high throughput DRAM-based TRNG can enable memory-centric system applications to source random values directly within the memory itself, thereby enhancing the overall potential, security, and privacy of such architectures.

Prior DRAM-based TRNG designs use DRAM data retention failures~\cite{keller2014dynamic, sutar2018d},
DRAM startup values~\cite{tehranipoor2016robust, eckert2017drng}, and non-determinism in DRAM command
scheduling~\cite{pyo2009dram} to generate true random numbers. Unfortunately, these approaches
do not fully satisfy these requirements because they either do not exploit a fundamentally
non-deterministic entropy source (e.g., DRAM command scheduling) or are too slow for continuous
high throughput operation (e.g., DRAM data retention failures, DRAM startup values).

Recent works~\cite{kim2019d,olgun2021quac,yuksel2025indram} enable high throughput, low-latency true random number generation in commodity off-the-shelf DRAM chips
by exploiting the failures induced by violating DRAM access timing parameters and activating multiple DRAM rows simultaneously. D-RaNGe~\cite{kim2019d}
reduces the DRAM row activation latency below manufacturer-recommended specifications to induce activation
failures and demonstrates that the resulting TRNG cells fail truly randomly. D-RaNGe~\cite{kim2019d} provides over two orders of magnitude higher throughput
than the prior state-of-the-art DRAM-based TRNG at that time. QUAC-TRNG~\cite{olgun2021quac} activates four DRAM
rows simultaneously, causing the bitline sense amplifiers to non-deterministically converge to
random values, and achieves lower latency and higher throughput than D-RaNGe. The state-of-the-art
DRAM-based TRNG, SiMRA-TRNG~\cite{yuksel2025indram}, generalizes quadruple-row activation~\cite{olgun2021quac} to simultaneous N-row activation~\cite{yuksel2024simultaneous}, where N can be up to 32.

\subsection{State-of-the-Art DRAM-based TRNG:\\SiMRA-TRNG}
\begin{figure}[t]
  \centering
 \includegraphics[width=1\linewidth]{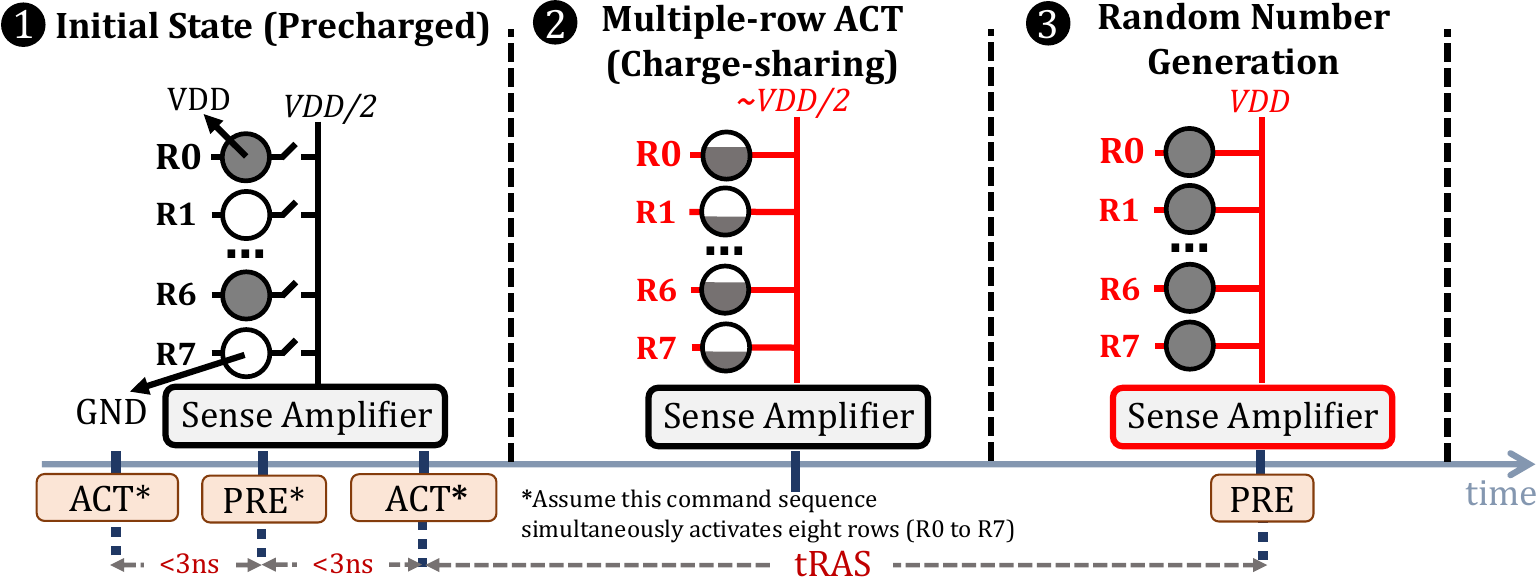}
  \caption{The command sequence for true random number generation using SiMRA. An
  ACT$\rightarrow$PRE$\rightarrow$ACT sequence simultaneously activates multiple rows; the opposing
  contributions of cells holding conflicting data perturb the bitline around $\sim$VDD/2, and the
  sense amplifier samples a random value.}
  \label{fig:simra-trng-mech}
\end{figure}

\noindent\textbf{Key Idea.} Fig.~\ref{fig:simra-trng-mech} shows the command sequence for true
random number generation using SiMRA, with eight DRAM cells connected to a bitline. Initially,
four cells have a voltage level of VDD, the remaining four are at ground (GND), and the bitline
has a voltage level of VDD/2. To generate random numbers, SiMRA-TRNG issues one
ACT$\rightarrow$PRE$\rightarrow$ACT command sequence with reduced DRAM timings. Doing so activates the cells simultaneously
and enables charge sharing between them and their bitline. Hence, the bitline voltage is perturbed by the simultaneously activated cells, where some try to pull up the
bitline voltage, and the others try to pull it down. As a result, these opposing contributions of
simultaneously activated cells can randomly perturb the bitline from the reference voltage. The sense amplifier then kicks in and tries to amplify the voltage on the bitline,
which results in sampling a random value based on the random perturbations on the bitline voltage.

\noindent\textbf{Design.} SiMRA-TRNG generates true random numbers using four banks in five
steps: 1)~selecting a high-entropy simultaneously activated row (SAR) group based on the characterization,
2)~initializing the rows that are used to generate true random numbers by performing the RowClone operation~\cite{seshadri2013rowclone,gao2019computedram,olgun2022pidram,yuksel2024simultaneous,yuksel2025pudhammer} in quick succession,
3)~performing ACT$\rightarrow$PRE$\rightarrow$ACT command sequence with reduced DRAM timings to activate the high-entropy SAR group, thereby generating random bits in the sense amplifiers of
four DRAM banks concurrently, 4)~reading from the four banks until 256 bits of Shannon entropy is
obtained, and 5)~post-processing the extracted bitstream using the SHA-256 cryptographic hash
function~\cite{fips2012180} to eliminate bias~\&~correlation.

\begin{figure}[t]
  \centering
  \includegraphics[width=\linewidth]{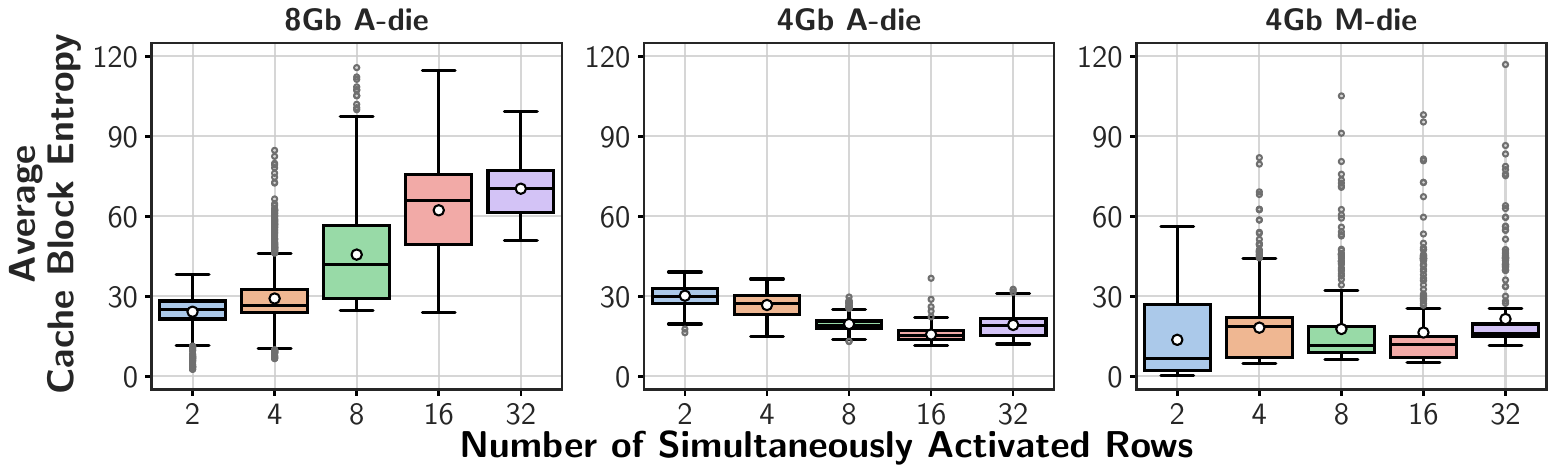}
  \caption{Average 512-bit cache-block entropy for varying numbers of simultaneously activated
  rows, across DRAM chip densities and die revisions.}
  \label{fig:simra-trng-entropy}
\end{figure}

\noindent\textbf{Real DRAM Chip Characterization.} We conduct a rigorous experimental characterization of 96 COTS DDR4 chips using an FPGA-based DRAM testing infrastructure~\cite{olgun2023dram,safari-drambender} (built on top of SoftMC~\cite{hassan2017softmc,softmcgithub}). We sweep the
number of simultaneously activated rows, the data pattern, temperature, and spatial location,
showing that SiMRA can generate random values with all the tested numbers of simultaneously activated
rows. We highlight three new empirical observations from this experiment. First, entropy tends to increase with the number of simultaneously
activated rows: across all tested chips, SiMRA generates 22.87, 25.17, 29.47, 34.54, and 40.41 average
cache-block entropy with 2-, 4-, 8-, 16-, and 32-row activation, respectively (Fig.~\ref{fig:simra-trng-entropy}).
Second, entropy also depends strongly on the data pattern, and is highest when the data pattern has an
equal number of logic-1s and logic-0s. We hypothesize this is because during charge sharing, logic-1 cells attempt to raise the bitline voltage while logic-0 cells attempt to lower it; thus, the bitline voltage converges to the reference voltage, the differential voltage falls well below the sense amplifier's reliable sensing margin, and the sense amplifier non-deterministically settles to logic-1 or logic-0.
Third, temperature affects the SiMRA's entropy. For example, for 32-row activation, as the temperature increases from 50\degC{} to 90\degC{}, average cache-block entropy decreases by 1.53\x{}.

\begin{figure}[t]
  \centering
\includegraphics[width=1\linewidth]{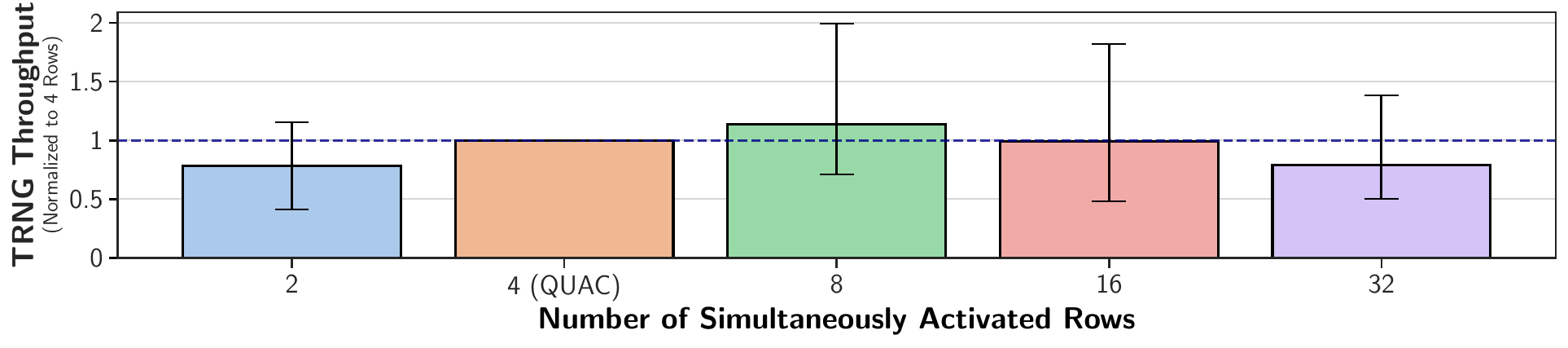}
  \caption{Throughput of generating true random numbers using SiMRA for varying numbers of
  simultaneously activated rows, normalized to the state-of-the-art DRAM-based TRNG (QUAC-TRNG,
  4-row activation)~\cite{olgun2021quac}.}
  \label{fig:simra-trng-tput}
\end{figure}

\noindent\textbf{Quality and Throughput.} True random numbers generated by SiMRA-TRNG pass \emph{all} NIST STS tests~\cite{rukhin2001statistical}, and SiMRA-TRNG reliably produces high-quality true random bitstreams for all tested numbers of simultaneously activated rows. SiMRA-TRNG designs based on 2-, 8-, 16-, and 32-row activation outperform the state-of-the-art DRAM-based TRNG (QUAC-TRNG, which uses 4-row activation), achieving up to
1.15\x{}, 1.99\x{}, 1.82\x{}, and 1.39\x{} higher throughput, respectively
(Fig.~\ref{fig:simra-trng-tput}). The 2-, 4-, 8-, 16-, and 32-row activation-based designs provide
average throughputs of 9.90, 13.81, 16.05, 13.94, and 10.81~Gb/s, respectively. There is a
tradeoff between TRNG latency and the entropy produced by SiMRA: while simultaneously activating
more rows tends to generate more entropy, it also increases the TRNG latency, so the higher
entropy of 32-row activation does not lead to higher throughput, because the latency of
initializing the simultaneously activated rows is doubled compared to 16-row activation.

SiMRA-TRNG outperforms the state-of-the-art DRAM-based TRNG in throughput and offers a promising
approach to generating true random numbers with high throughput, directly inside real, unmodified
COTS DRAM chips.

\subsection{In-DRAM Physical Unclonable Functions}

A physically unclonable function (PUF) maps a set of input parameters to unique,
device-specific signatures that can be generated repeatably and reliably. The resulting
signature reflects a device's inherent, random physical variations introduced during
manufacturing, which makes it practically impossible to predict or replicate without access to the device itself~\cite{gassend2002silicon,yan2015novel}. These characteristics enable PUFs to be frequently used in security
applications, such as low-cost authentication against security attacks and prevention of
integrated-circuit counterfeiting, typically through a challenge-response protocol in which a trusted server challenges a device and verifies its PUF response~\cite{sutar2018d,xiong2016run}. A runtime-accessible PUF, i.e., one that an application can evaluate on demand during online operation, must additionally be evaluated with low latency and cause low system interference.

DRAM-based PUFs are attractive for two key reasons. First, DRAM is already widely used in modern systems, ranging from embedded to server. Second, DRAM's large address space provides a greater challenge-response space than smaller components such as SRAM. Prior DRAM PUF proposals exploit variations in DRAM start-up values, DRAM access latencies, and DRAM cell retention failures, and DRAM read disturbance to generate PUF responses~\cite{keller2014dynamic,sutar2016d,sutar2018d,tehranipoor2016robust,xiong2016run,schaller2019decay,liu2014trustworthy,zheng2021implementation,kumari2018rapid,mexis2025achieving,talukder2019prelatpuf,hashemian2015robust,najafi2021deep,najafi2025epuf,miskelly2020fast}. Unfortunately, these proposals are unsuitable as runtime-accessible PUFs: start-up-value PUFs require a DRAM power cycle for every authentication, write-latency PUFs require additional circuitry in the DRAM chip, and retention-failure PUFs are too slow, taking on the order of minutes to evaluate at typical operating temperatures.

Recent works~\cite{kim2018dram,baser2026simrapuf,orosa2021codic,gao2022frac} enable fast, reliable, runtime-accessible PUFs
even in modified and commodity off-the-shelf DRAM chips by exploiting the error patterns induced by violating DRAM
access timing parameters. The DRAM Latency PUF~\cite{kim2018dram} is a fast, reliable, runtime-accessible DRAM-based PUF whose key
idea is to reduce the DRAM read access latency below the reliable manufacturer-recommended specifications. Doing so results in error patterns that reflect the compound effects
of manufacturing variations in various DRAM structures (e.g., capacitors, transistors, sense
amplifiers): some DRAM cells fail always after repeated accesses with violated timing parameters
and some others never fail at all, and a combination of such consistently failing or never failing
cells generates a unique identifier for the device. An experimental characterization of 223 LPDDR4
chips from all three major manufacturers shows that these error patterns are quickly generated (at
88.2~ms) irrespective of operating temperature, without any modification to the DRAM chip.
Another line of work generates signatures by controlling DRAM's internal circuitry and sensing
cells at fractional voltage levels. CODIC~\cite{orosa2021codic} is a low-cost DRAM substrate
that enables fine-grained control over four previously fixed internal DRAM signals, including
those that trigger the sense amplifiers; by controlling when the sense amplifiers fire, CODIC
senses cells at fractional voltage levels, and its CODIC-sig command generates digital signatures
that depend on process variation. A CODIC-based PUF that uses these signatures provides 1.8\x{}
higher throughput than the best prior DRAM-based PUF at that time, with similar resilience to temperature changes and
more repeatable responses. The Frac-based PUF~\cite{gao2022frac}, the prior state-of-the-art
DRAM-based PUF, builds on top of CODIC to leverage fractional voltage levels on commodity
off-the-shelf chips without any modification: it initializes a target row with all ones and
repeatedly applies the \texttt{Frac} operation (back-to-back ACT$\rightarrow$PRE command pairs with
reduced timing) to drive the DRAM cells toward $V_{DD}/2$, after which a subsequent activation
resolves each cell to logic-0 or logic-1 according to the manufacturing process variation.
The current state-of-the-art DRAM-based PUF, SiMRA-PUF~\cite{baser2026simrapuf}, leverages simultaneous multiple-row activation to generate device-specific signatures and outperforms the Frac-based PUF in evaluation latency, which we describe in detail next.

\subsection{State-of-the-Art DRAM-based PUF:\\SiMRA-PUF}

\noindent\textbf{Key Idea.} The \emph{key idea} of SiMRA-PUF is to generate device-specific
signatures by exploiting the charge-sharing process that arises from simultaneously activating
multiple DRAM rows holding opposing data patterns. An ACT$\rightarrow$PRE$\rightarrow$ACT (APA) command sequence with reduced timings simultaneously
activates multiple rows initialized with a balanced data pattern, enabling charge sharing between
the activated cells and the bitlines. Because the cells hold opposing values, the resulting
bitline perturbation does not strongly deviate from the reference voltage. However, due to
process- and design-induced variations, each bitline settles at a voltage near, but slightly
different from, the reference voltage. The sense amplifiers then resolve the bitlines to logic-0
or logic-1: for some bitlines, the perturbation falls below the reliable sensing margin and the
sense amplifier resolves the bitline randomly across SiMRA trials, whereas for others the
perturbation exceeds the margin and the sense amplifier resolves the bitline to a single, stable
value across trials. These stable, device-specific bits form the signature.

\noindent\textbf{Challenge and Response.} SiMRA-PUF defines each challenge as a (bank, subarray)
pair. To generate a response, it initializes a simultaneously-activated-row (SAR) group with a
balanced data pattern, issues an APA command sequence, and reads the sense amplifier outputs as
the signature; the PUF response is the signature read from the selected SAR group.

\begin{figure}[t]
  \centering
  \includegraphics[width=0.75\linewidth]{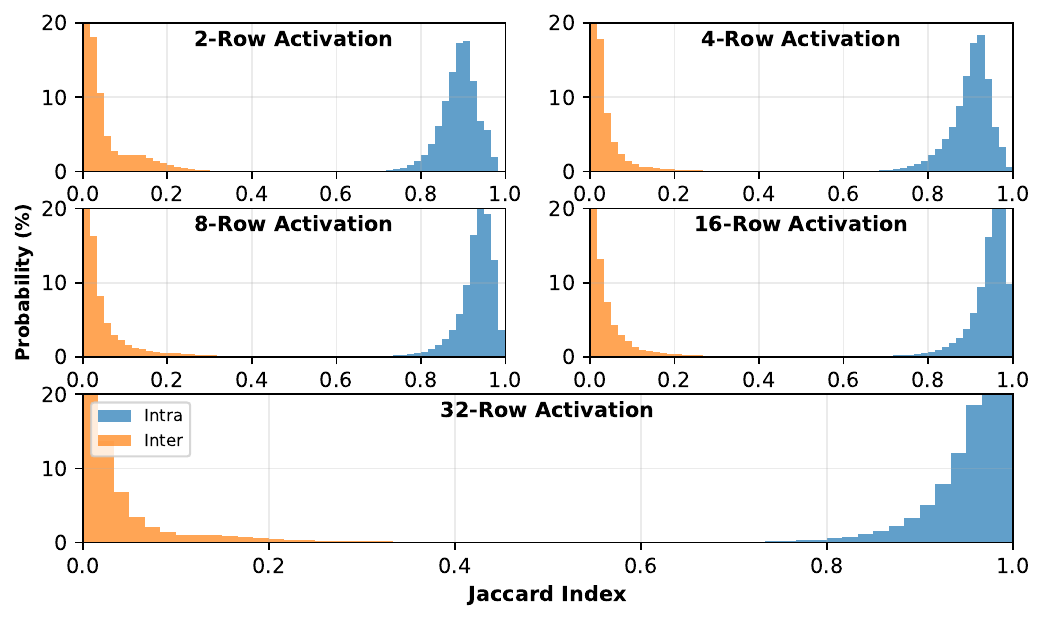}
    \caption{Inter- (orange) and intra-Jaccard (blue) indices obtained for 2/4/8/16/32-row activation-based SiMRA-PUF.}  \label{fig:simra-puf}
\end{figure}

\noindent\textbf{Real DRAM Chip Characterization.} SiMRA-PUF is characterized on 112 COTS DDR4
chips (from 10 modules), using the intra-Jaccard index (response stability, higher is better) and
the inter-Jaccard index (response uniqueness, lower is better). SiMRA-PUF provides average
inter-Jaccard indices of 3.98\%, 2.37\%, 3.44\%, 2.92\%, and 3.24\%, and average intra-Jaccard
indices of 89.02\%, 89.81\%, 93.03\%, 94.06\%, and 94.86\%, for 2-, 4-, 8-, 16-, and 32-row
activation, respectively (Fig.~\ref{fig:simra-puf}). SiMRA-PUF thus produces unique and repeatable
responses across all tested activation counts. 

\noindent\textbf{Comparison with the Prior State-of-the-Art.}  SiMRA-PUF provides intra-Jaccard indices similar to the Frac-based PUF~\cite{gao2022frac} (95.62\%), with the gap narrowing as the number of simultaneously activated rows increases. Most importantly, 2-row activation-based SiMRA-PUF has 5.75\% lower evaluation latency than the Frac-based PUF, although evaluation latency grows with the number of activated rows. For example, 32-row activation reaches 3.84\x{} the latency of Frac-based PUF~\cite{gao2022frac}.

These results show that SiMRA generates device-specific signatures suitable for high-quality, reliable PUF responses, and that 2-row SiMRA-PUF achieves lower evaluation latency than the prior state-of-the-art DRAM-based PUF, directly on real, unmodified COTS DRAM chips.
\section{Security Challenges in Processing-in-DRAM}

As a new processing paradigm, Processing-in-DRAM introduces new security considerations
related to its integration in real-world computing systems. We focus on two of them, one
affecting reliability and one affecting confidentiality.

First, modern DRAM is subject to
read disturbance, a worsening safety, security, and reliability phenomenon in which repeatedly accessing one DRAM row induces bitflips in physically nearby, unaccessed rows~\cite{kim2020revisiting,orosa2021deeper,olgun2023dram,olgun2024read,olgun2025variable,yuksel2025columndisturb,luo2023rowpress,luo2024rowpress,kim2014flipping,mutlu2017rowhammer,mutlu2019rowhammer,mutlu2023fundamentally,yuksel2025pudhammer,yaglikci2022understanding,lim2017active, park2016statistical, park2016experiments, ryu2017overcoming, yun2018study, lim2018study,mutlu2025memory,lang2023blaster,nam2024dramscope, olgun2023hbm,nam2023xray,luo2025revisiting,he2023whistleblower,luo2024experimental,hassan2021utrr,frigo2020trrespass,wang2026scaledisturb,luo2026dejavu}.
RowHammer~\cite{kim2014flipping} and RowPress~\cite{luo2023rowpress} are two prominent examples,
where a row (victim row) experiences bitflips when a nearby row (aggressor row) is repeatedly activated
(hammered) or kept open for a long period (pressed). Since PuD operations rely on multiple-row
activation, where multiple DRAM rows are activated simultaneously or in quick succession, they
can exacerbate DRAM read disturbance, as PuDHammer~\cite{yuksel2025pudhammer}
shows.

Second, the adoption of processing-in-memory provides applications with a new, direct, and fast way to access main memory, which malicious applications can exploit. Naively providing such access may lead to unforeseen information leakage and other confidentiality and integrity issues. In particular, it opens high throughput timing attack vectors that are hard to mitigate without significant performance overhead, as IMPACT~\cite{bostanci2025revisiting} shows.

\subsection{Read Disturbance Effects\\of Processing-using-DRAM: PuDHammer}

PuDHammer~\cite{yuksel2025pudhammer} presents the first experimental characterization of the
read disturbance effects of Processing-using-DRAM (PuD) operations, on 316 real DDR4 chips
from four major manufacturers.

\noindent\textbf{Key Idea.} PuD operations leverage an analog DRAM operation called multiple-row
activation, where multiple DRAM rows are activated simultaneously or in quick succession.
PuDHammer characterizes the read-disturbance effects of its two types: consecutive multiple-row
activation (CoMRA), used for in-DRAM data copy \& initialization, and simultaneous multiple-row activation (SiMRA),
used for in-DRAM bitwise operations. To hammer with CoMRA, PuDHammer repeatedly performs
consecutive activation of a source and a destination row; to hammer with SiMRA, it issues an
ACT$\rightarrow$PRE$\rightarrow$ACT command sequence with reduced DRAM timings to simultaneously activate
multiple rows. In both cases, the activated rows become aggressor rows and their neighboring rows
become victim rows; when two activated rows sandwich a victim row, this forms a double-sided
attack.

\begin{figure}[ht]
    \centering
    \includegraphics[width=1\linewidth]{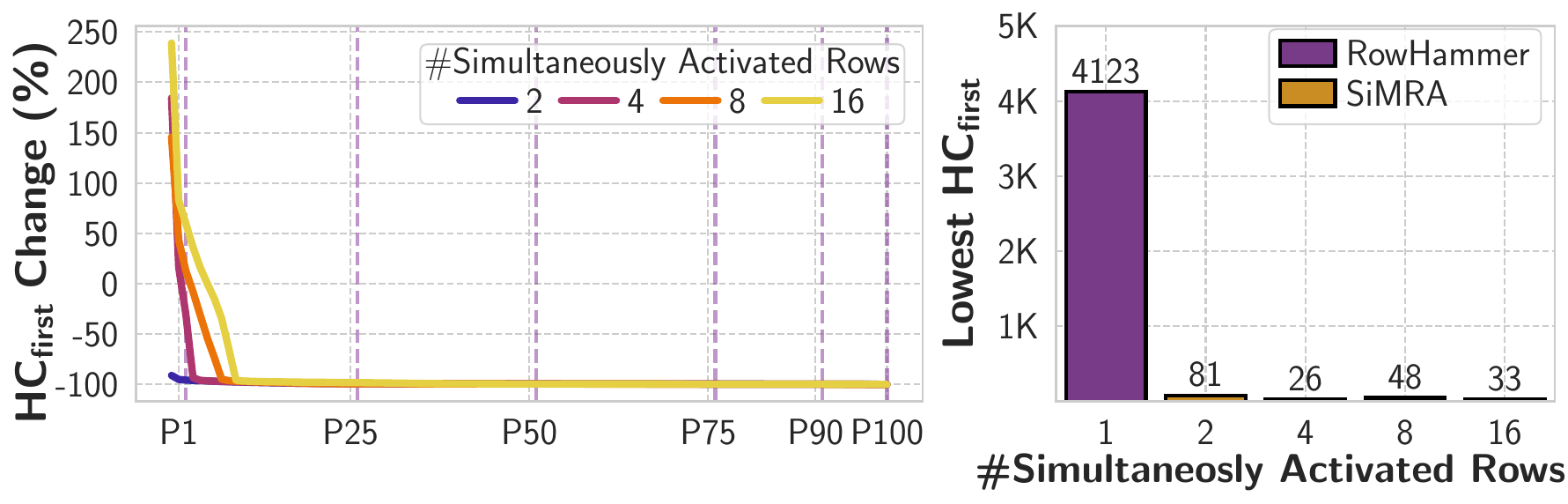}
    \caption{Distribution of the change in $HC_{first}$ change with double-sided SiMRA compared to double-sided RowHammer (left) and the lowest $HC_{first}$ observed with double-sided SiMRA and RowHammer (right).}
    \label{fig:simra_rh_ds}
\end{figure}

\noindent\textbf{Takeaway from Real DRAM Chip Characterization.} Both CoMRA and SiMRA greatly increase a DRAM chip's read-disturbance
vulnerability, decreasing the minimum hammer count required to induce the first bitflip ($HC_{first}$)
in all tested DRAM chips from four manufacturers. The effect is strongest for SiMRA: across all tested
chips, the lowest $HC_{first}$ observed with double-sided SiMRA is up to 158.58\x{} lower than with
double-sided RowHammer (Fig.~\ref{fig:simra_rh_ds}), while the lowest $HC_{first}$ with double-sided
CoMRA is up to 13.98\x{} lower than with double-sided RowHammer. 
PuD operations thus make a DRAM chip far more vulnerable to read disturbance than conventional read disturbance access patterns (e.g., RowHammer).

\noindent\textbf{Mitigation.} PuDHammer adapts and evaluates the industry's state-of-the-art
RowHammer mitigation, Per-Row Activation Counting (PRAC)~\cite{jedec2024jesd795c}, and finds that the adapted PRAC solution
incurs an average system performance overhead of 48.26\% across 60 five-core multiprogrammed
workloads, showing that extending existing RowHammer defenses to PuD is costly and that
read-disturbance-resilient PuD systems call for new solutions.

\subsection{Timing Attacks in Processing-in-DRAM:\\IMPACT}

IMPACT~\cite{bostanci2025revisiting} is a set of high throughput main memory-based timing attacks that
leverage characteristics of processing-in-memory (PiM) architectures to establish covert and side
channels.

\noindent\textbf{Key Idea.} The adoption of PiM provides userspace applications with fast and
reliable direct access to main memory that bypasses the cache hierarchy. IMPACT observes and
exploits the shared DRAM row buffer: because a row-buffer conflict takes measurably longer than a
row-buffer hit, an application can infer whether a row was recently accessed by timing its
accesses. IMPACT achieves high throughput by (i)~eliminating the expensive cache-bypassing steps
required by processor-centric timing attacks and (ii)~leveraging the intrinsic parallelism of PiM
operations.

\noindent\textbf{Attacks.} IMPACT is demonstrated with three case studies. First, IMPACT-PnM is a covert channel that
exploits PiM-enabled instructions~\cite{PEI}, a PnM mechanism that executes simple operations in compute
units near DRAM: the sender encodes a message as row-buffer conflicts at selected addresses, and
the receiver detects them by timing how long a PiM-enabled instruction takes to operate on those
rows. Second, IMPACT-PuM is a covert channel that instead exploits RowClone~\cite{seshadri2013rowclone}, a PuM operation that performs
bulk in-DRAM row copy: the sender transmits bits by copying rows in different DRAM banks in
parallel, and the receiver decodes them by issuing its own row-copy operations and measuring their
latency. Third, IMPACT's side-channel attack uses PiM-enabled instructions to leak the private
information of a concurrently-running application by observing its memory access patterns. 
IMPACT side channel attack is demonstrated on a genomic privacy proof-of-concept attack, where the attacker leaks private characteristics of a user’s query genome by observing the
memory access patterns of a read mapping application.

\begin{figure}[t]
  \centering
  \includegraphics[width=1\linewidth]{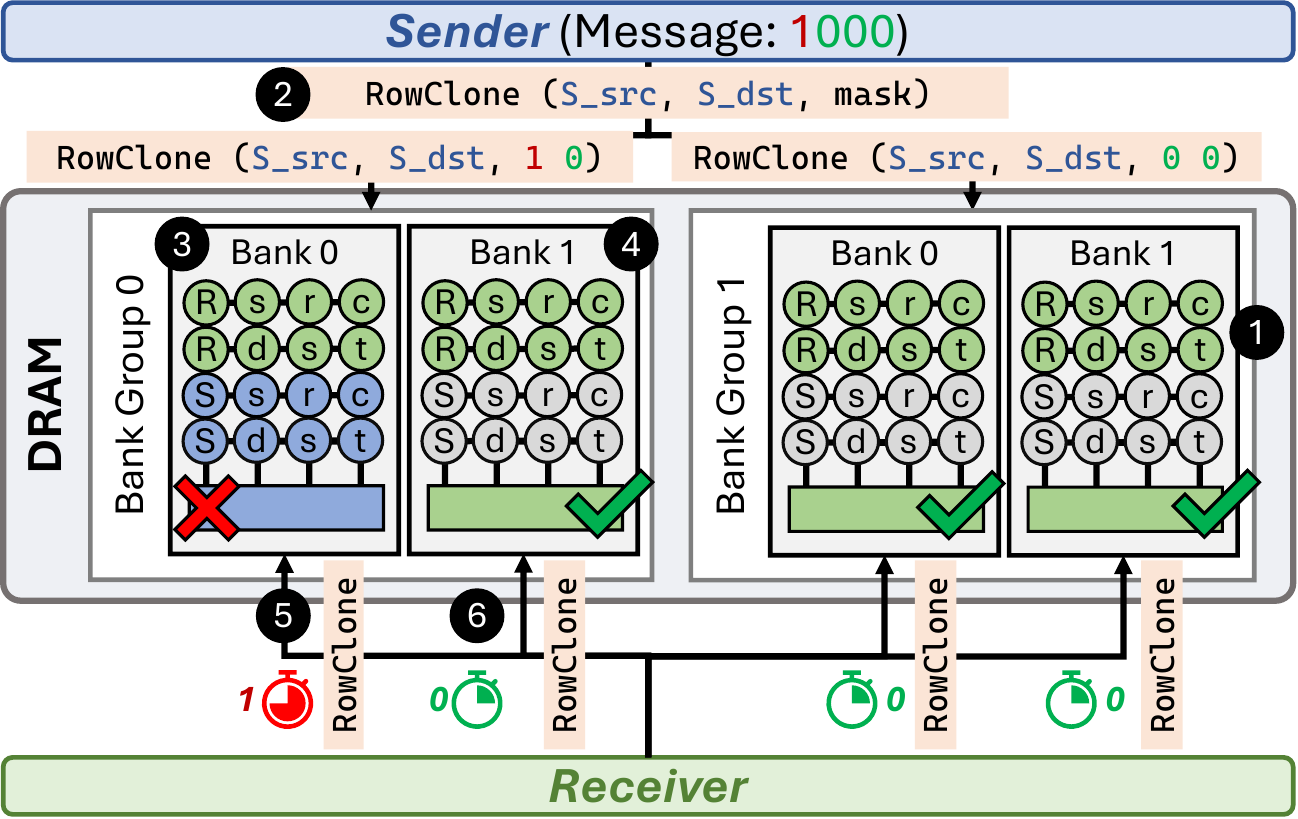}
  \caption{PuM covert-channel attack flow (IMPACT-PuM).}
  \label{fig:impact-pum}
\end{figure}

\noindent\textbf{Example: the IMPACT-PuM Covert Channel.} We use IMPACT-PuM to illustrate IMPACT's
end-to-end flow (Fig.~\ref{fig:impact-pum}). Before the attack, the sender and receiver co-locate
their data in the same set of DRAM banks. The sender then transmits the message in $M$-bit
batches: it maps each bit of a batch to a separate DRAM bank and builds a mask, so that a single
RowClone operation transmits an entire $M$-bit batch in parallel (where $M$ is the number of
banks). To send a 1, the sender triggers RowClone in the corresponding bank, activating a
different row from the receiver's and causing interference; to send a 0, it issues no RowClone and
does not interfere. The receiver decodes the batch by issuing one RowClone operation per bank and
measuring its latency: a high latency indicates a row-buffer conflict (logic-1), and a low latency
indicates a row-buffer hit (logic-0).

\noindent\textbf{Result.} IMPACT-PuM achieves a communication throughput of 14.8~Mb/s, 6.5\x{}
higher than the state-of-the-art main memory-based covert channel, and the side-channel attack on
genomic read mapping leaks the private characteristics of a user's query genome at 7.6~Mb/s with
96\% accuracy.

\noindent\textbf{Mitigation.} IMPACT evaluates four defense mechanisms that attempt to eliminate
the timing channel, but finds that mitigating IMPACT incurs high performance overheads, concluding that more
research is needed to find low-overhead solutions.
\section{Conclusion and Outlook}

We examined the security benefits and challenges of memory-centric computing, specifically
Processing-in-DRAM (PiD). 
By eliminating data movement between memory
and processor, the PiD paradigm takes a large step towards getting rid of one of the most attacker-exposed types of data movement within a computing node, i.e., data movement over the main memory bus. Enabling the secure and private execution of computations in PiD
systems can therefore potentially enable fundamentally more secure computing systems. This requires providing support for such secure computation. For example, our afore-described SiMRA-TRNG and SiMRA-PUF are two notable examples of novel in-DRAM security primitives that take advantage of PiD.

As in existing systems, robustness and information security are important in PiD systems, where DRAM rows can be frequently activated and deactivated and/or can be accessed directly via a fast path. We described two such works that can amplify the current robustness and security issues in computing systems: PuDHammer, which greatly exacerbates DRAM read disturbance, and IMPACT, which exploits direct, fast memory access to amplify high throughput timing attacks.

We firmly believe that it is time to architect principled computing systems where DRAM (and in general, memory; see~\cite{mutlu2020modern,ghose.ibmjrd19,mutlu2019processing} for a broad discussion) is designed, used, and programmed not as an inactive storage substrate, but instead as a \emph{combined computation, storage, and security substrate} where computational capability, storage density, and security are key goals. Although many challenges remain to enable widespread adoption of Processing-in-DRAM (and memory-centric systems in general), we believe the mindset and infrastructure shift necessary to enable such a secure computation-storage paradigm remains to be the largest challenge. Overcoming this mindset and infrastructure shift can unleash a fundamentally more secure, energy-efficient, and high-performance way to design, use, and program computing systems.

We therefore believe the future of memory-centric systems is very bright and promising, yet there needs to be many exciting challenges to be solved across the computing stack to facilitate widespread and easy adoption.

\section*{Acknowledgments}
We thank the SAFARI Research Group members for providing a stimulating intellectual and scientific environment. We acknowledge the generous gifts from our industrial partners, including Google, Huawei, Intel, and Microsoft. This work, along with our broader work in Processing-in-Memory and memory systems~\cite{kim2014flipping,mutlu2020modern,mutlu2013memory,mutlu2025memory,mutlu2019rowhammer,mutlu2023fundamentally,mutlu2019processing,cai2017flashtbd,mutlu2014research,singh2021fpga,mutlu2017rowhammer,mutlu2020intelligentdate,oliveira2022accelerating,mutlu2024memory,mutlu2023experimentalretrospective,mutlu2023raidrretrospective,mutlu2023rowhammerretrospective, ahn2023retrospective, mutlu2023selfretrospective,luo2026ramulator2.1,bostanci2026extended,olgun2026drambender,deoliviera2021damov,mutlu2015main,cai2017errors,gomez2021benchmarking,kakolyris2026columnkeeper,ghose.ibmjrd19}, is supported in part by the Semiconductor Research Corporation (SRC), the ETH Future Computing Laboratory (EFCL), ACCESS – AI Chip Center for Emerging Smart Systems, a Google Security and Privacy Research Award, and the Microsoft Swiss Joint Research Center.

\balance
\bibliographystyle{unsrt}
\bibliography{refs}

\end{document}